\documentstyle[pra,aps,epsf,preprint]{revtex}
\tightenlines

\begin{document}
\title{The role of symmetry on interface states in magnetic tunnel junctions}
\author{C. Uiberacker and P. M. Levy}
\address{Department of Physics, New York University,\\
4 Washington Place, New York, NY 10003 }
\date{\today}
\maketitle

\begin{abstract}
When an electron tunnels from a metal into the barrier in a magnetic tunnel
junction it has to cross the interface. Deep in the metal the eigenstates
for the electron can be labelled by the point symmetry group of the bulk but
around the interface this symmetry is reduced and one has to use linear 
combinations of the bulk states to form the eigenstates
labelled by the irreducible representations of the point symmetry group of
the interface. In this way there can be states localized at the interface 
which control tunneling. The conclusions as to which are the dominant 
tunneling states are different from that conventionally found.
\end{abstract}

%\twocolumn[\hsize\textwidth\columnwidth\hsize\csname @twocolumnfalse\endcsname
%\draft
%\widetext 

%\draft

\begin{figure}[h] 
\begin{center}
\leavevmode
\begin{tabular}{cc}
 \begin{minipage}[t]{7cm}
   \epsfxsize=7cm   
%  \epsffile{/home/cu2/plots/Fe/DOS/radial/delV_e-9/dos_av1SPD_dn.ps}
   \epsffile{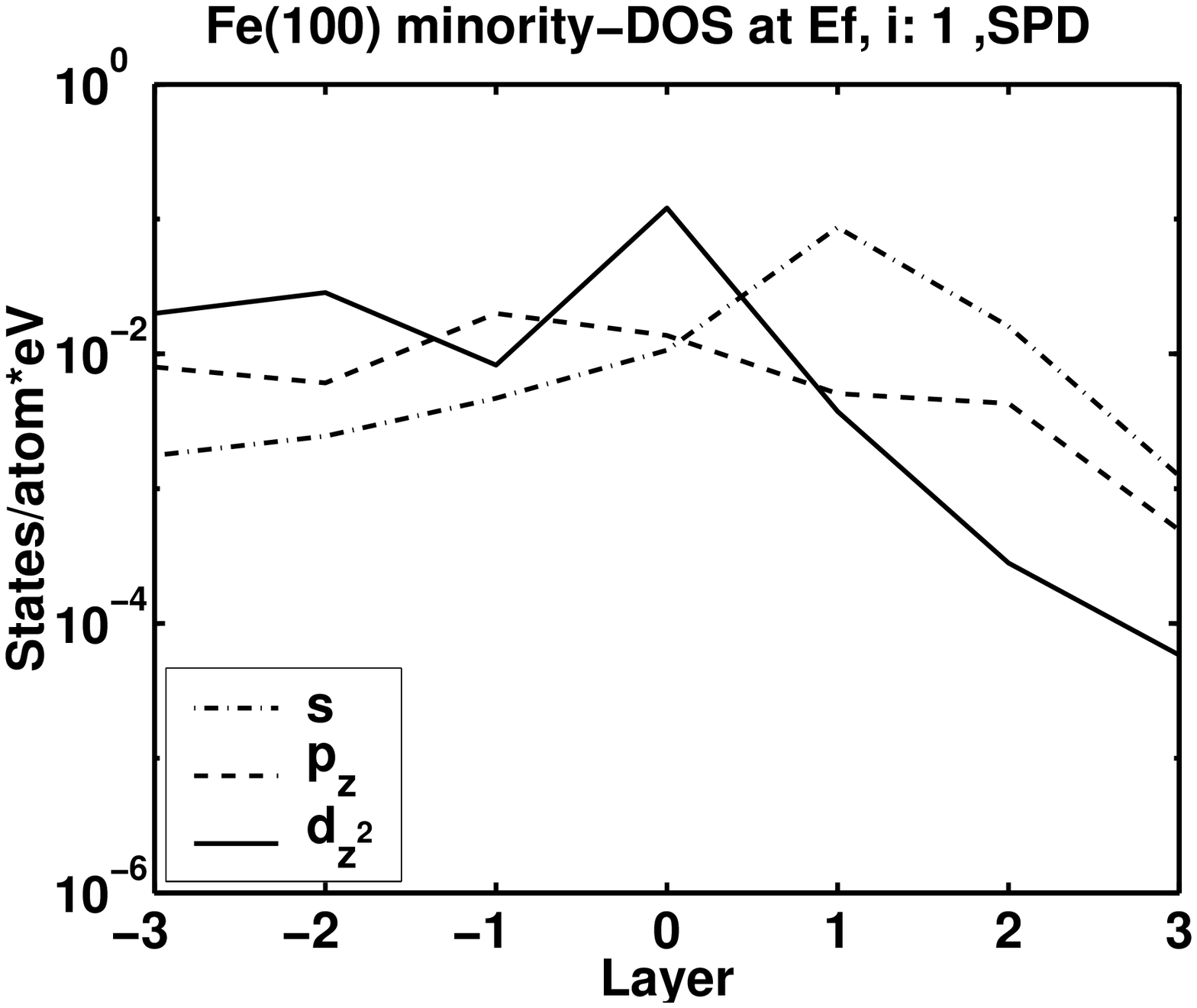}
 \end{minipage} &
 \begin{minipage}[t]{7cm}
   \epsfxsize=7cm
%  \epsffile{/home/cu2/plots/Fe/DOS/radial/delV_e-9/dos_av1HYB_dn.ps}
   \epsffile{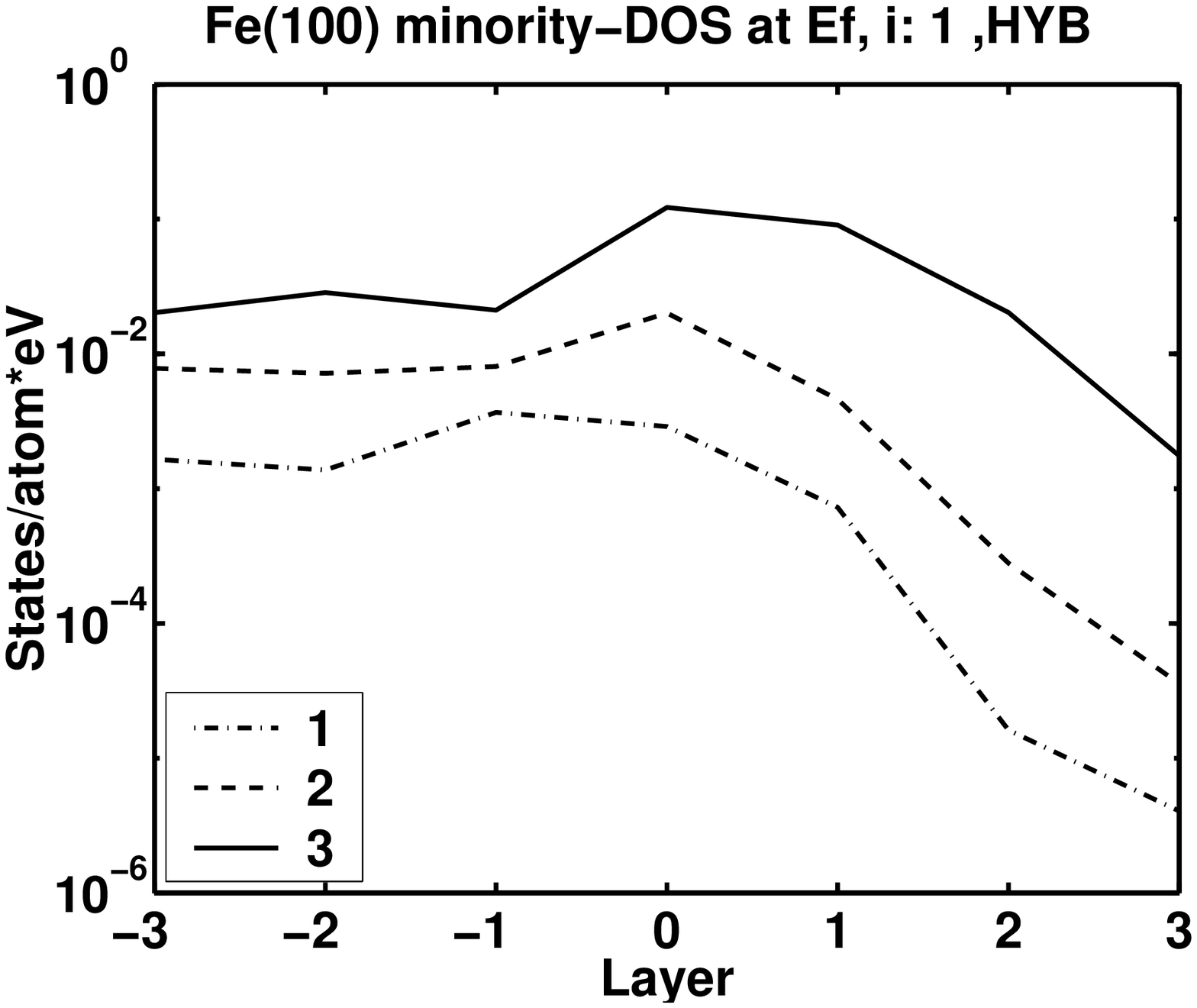}
 \end{minipage} 
 \end{tabular}
\caption{$\Delta_1$ contribution to the integrated DOS at the Fermi 
Energy plotted on each layer in the IR basis (left) and eigenbasis 
(right) for minority states. Layer $0$ denotes the surface layer 
and layers$>$0 denote vacuum.}
\label{dos_prof}
\end{center} 
\end{figure}

%\phantom{.}
%] \narrowtext
\newpage Magnetic tunnel junctions (MTJ's) are layered structures of the
form ferromagnetic metal/insulator/ferromagnetic metal, and provide a very
interesting area of basic research. The tunneling currents of MTJ's are very
much influenced by the electronic structure around the interface between the
electrodes and the insulating barrier. For example by inserting a few layers
of Cu (dusting layers) between one of the ferromagnetic Co electrodes and
the Al$_{2}$O$_{3}$ barrier it was found recently \cite{leclair1,leclair2}
that the TMR ratio is suppressed exponentially. Furthermore, by putting 1ML
Cr between Co and the same barrier the TMR effect nearly vanishes but when
inserting 3-5 ML of Co between the Cr layer and the barrier the TMR was
restored again \cite{leclair3}. There have been theoretical attempts to
model MTJ's with such dusting layers by using simple semiclassical models %
\cite{zhang1} and tight-binding schemes \cite{mathon1} that gave some
insight on the role of Fermi surface mismatch between different metals and
resulting quantum well states by defining a tunneling density of states.
When quantum mechanics is used to study electrons in crystals it follows
from the symmetry of the Hamiltonian that the energies and eigenstates can
be labelled by the irreducible representations (IR's) of the space group.
Because of the strong screening of the Fermi sea of a metal electrons are
not aware of the interface until they are within 3-4 monolayers of it;
therefore electron states inside a metal far from the interface are
labelled by the IR's of the point group in the bulk.  Close to the 
interface the electron states should be classified according to the IR's 
of the point group of the interface which is a subgroup of the point group 
in the bulk. In addition, states that are localized at the surface may form,
e.g., on transition-metals with (100) surfaces \cite{stroscio1}.  These two
features can be seen in Fig.(\ref{dos_prof}) by comparing the density of
states for minority electrons using as a basis the bulk eigenfunctions 
$d_{z^{2}},p_{z},s$ (left figure) and the true eigenfunctions (right figure)
at an Fe(100)/vacuum interface. Three layers away from the interface the
true eigenfunctions are nearly identical to the bulk eigenfunctions, but in
a range of 2 layers around the interface $d_{z^{2}},p_{z},s$ cannot be used
to describe the electron; rather one has to use appropriate linear
combinations of these states to form the true eigenstates. 
We also note in these figures that 
the DOS peaks in the region of the interface and decay exponentially into 
the barrier but also decay inside the metal indicating that the minority 
states are localized at the interface.
In this brief report we outline the procedure for identifying which
states are primarily responsible for tunneling in MTJ's.

In order to analyze the electronic states in a MTJ one calculates the
density of states (DOS) for it. Within a Green's function formalism and a
single particle picture the DOS operator can be written as 
\begin{equation}
\hat{\rho}(E;\delta )=\frac{1}{2\pi i}(\hat{G}^{A}-\hat{G}^{R})=\frac{1}{%
2\pi i}\left( \hat{G}(E-i\delta )-\hat{G}(E+i\delta )\right)   \label{dosop}
\end{equation}%
by making use of the operators $\hat{G}^{R(A)}$ corresponding to the
retarded and advanced Green's function. Here, $i$ denotes the complex unit
and the resolvent $\hat{G}(z)$ is defined by the equation 
\begin{equation}
(z-\hat{H})\hat{G}(z)=\hat{I}
\end{equation}%
with $z$ being a complex number, $\hat{H}$ being the Hamiltonian and $\hat{I}
$ being the unit operator. Using the eigenfunctions $|\Phi _{j}\rangle $ of
the Hamiltonian one can write the resolvent in the spectral representation
and the DOS operator becomes 
\begin{equation}
\hat{\rho}(E;\delta ) =\frac{1}{\pi }\sum_{j}|\Phi _{j}\rangle \langle \Phi _{j}|
\frac{\delta }{(E-E_{j})^{2}+\delta ^{2}};\quad \delta >0\quad .
\end{equation}%
Using the properties of $|\Phi _{j}\rangle \langle \Phi _{j}|$ 
it follows that $\hat{\rho}(E;\delta )$ is normal and therefore has real
positive definite diagonal elements in any basis. Especially, in a real space
representation expressed in an angular momentum basis the matrix elements     
\begin{equation}
\rho ({\bf r};E;\delta )=\sum_{LL^{\prime }}\left( \frac{1}{\pi }%
\sum_{j}\Phi _{j,L}(r)\Phi _{j,L^{\prime }}^{\ast }(r)\frac{\delta }{%
(E-E_{j})^{2}+\delta ^{2}}\right) Y_{L}({\bf \hat{r}})Y_{L^{\prime }}^{\ast
}({\bf \hat{r}})
\end{equation}%
are real and positive for each value of $r$.

The MTJ we choose to investigate is Fe(100)/Vac/Fe(100) with an underlying
b.c.c.\ lattice and we concentrated on the DOS around the surface in
Fe(100)/Vac. For calculating the DOS we used the {\it ab-initio}
spin-polarized scalar-relativistic Korringa-Kohn-Rostoker (KKR) multiple
scattering formalism for layered systems~\cite{bi289} together with the
local spin density approximation (LSDA) and the Gunnarsson-Lundqvist
exchange-correlation potential~\cite{bi283}. To obtain short ranged
structure constants, a screening formalism with a screening potential of $2.0
$~Ry was applied~\cite{bi289,bi299} and self-consistent potentials were
evaluated within the atomic sphere approximation (ASA). First we calculated
the potentials and exchange fields, where we used 16 points along a
semi-circle in the complex energy plane with the points distributed on a
Gaussian mesh and 45~$k_{||}$ points in the irreducible surface Brillouin
zone (ISBZ). The vacuum was simulated by six layers of empty atomic spheres
which have an underlying b.c.c.\ lattice structure. For calculating the DOS
we choose an imaginary part of $\delta=5\times 10^{-4}$ Ry and took 1830 $%
k_{||}$ points in the ISBZ for the minority and majority spin-channel.

In KKR for layered systems the arrangement of atoms within each layer is
decomposed into Wigner-Seitz cells. An arbitrary point ${\bf r}_n^p$ is then
uniquely described by the origin of cell $n$ in layer $p$, ${\bf R}_n^p$,
and the position ${\bf r}$ inside this cell and therefore ${\bf r}_n^p={\bf R%
}_n^p+{\bf r}$. In the ASA each cell is then replaced by a sphere of equal
volume and the potential inside the cell is assumed to be spherically
symmetric. It follows that the eigenfunctions of the Hamiltonian inside a
given cell can be written as a series in angular momentum with expansion
coefficients depending only on the distance from the origin of the cell.
Then the Green's function can be written as 
\begin{eqnarray}
G({\bf r}_n^p,{\bf r}_m^{^{\prime}q};E+i\delta) &=&
\sum_{LL^{\prime}}\left\{
Z_l^{n;p}(r)\tau_{LL^{\prime}}^{nm;pq}Z_{l^{\prime}}^{m;q}(r^{\prime})-%
\delta_{nm}\delta_{pq}\delta_{LL^{\prime}} \left[Z_l^{n;p}(r)J_l^{n;p}(r^{%
\prime})\theta(r^{\prime}-r) \right. \right.  \nonumber \\
&+&\left.\left. J_l^{n;p}(r)Z_l^{n;p}(r^{\prime})\theta(r-r^{\prime})\right]
\right\} Y_L({\bf \hat{r}})Y_{L^{\prime}}^*({\bf \hat{r}}^{\prime}) \quad ,
\end{eqnarray}
by using $L=(l,m)$ as a shorthand notation, in terms of the scattering path
operator $\tau_{LL^{\prime}}^{nm;pq}$ and the regular solutions $Z_l^{n;p}(r)
$ and the irregular solution $J_l^{n;p}(r)$ of the Schr\"odinger equation
which all depend implicitly on $E+i\delta$ \cite{bi289,bi299}. Using the
fact that the Hamiltonian is hermitean it immediately follows that $\hat{G}%
(E-i\delta)=\hat{G}^\dagger(E+i\delta)$ and inserting this into Eqn.(\ref%
{dosop}) yields 
\begin{eqnarray}
\rho({\bf r}_n^p,{\bf r}_m^{^{\prime}q};E;\delta) & = &\frac{1}{2\pi i}
\left\{\left[G({\bf r}_m^{^{\prime}q},{\bf r}_n^p;E+i\delta)\right]^* - G(%
{\bf r}_n^p,{\bf r}_m^{^{\prime}q};E+i\delta)\right\}  \nonumber \\
&=&\frac{1}{2\pi i}\sum_{LL^{\prime}}\left\{\left[G_{L^{%
\prime}L}^{mn;qp}(r^{\prime},r)\right]^*
-G_{LL^{\prime}}^{nm;pq}(r,r^{\prime})\right\} Y_L({\bf \hat{r}}%
)Y_{L^{\prime}}^*({\bf \hat{r}}^{\prime})
\end{eqnarray}
and the diagonal matrix elements are            
\begin{eqnarray}
\rho({\bf r}_n^p;E;\delta)&=&\sum_{LL^{\prime}}\rho_{LL^{\prime}}^{n;p}(r;E;%
\delta)Y_L({\bf \hat{r}}) Y_{L^{\prime}}^*({\bf \hat{r}})  \nonumber \\
&=&\sum_{LL^{\prime}}\left\{\left[Z_{l^{\prime}}^{n;p}(r)^*{%
\tau_{L^{\prime}L}^{nn;pp}}^*
Z_l^{n;p}(r)^*-Z_l^{n;p}(r)\tau_{LL^{\prime}}^{nn;pp}Z_{l^{\prime}}^{n;p}(r)%
\right]/2\pi i \right.  \nonumber \\
&+&\left. {\rm Im}\left[Z_l^{n;p}(r)J_l^{n;p}(r)\right]\right\}Y_L({\bf \hat{%
r}}) Y_{L^{\prime}}^*({\bf \hat{r}})  \label{rhoKKR}
\end{eqnarray}
It is very common to calculate the $L$-resolved integrated DOS (simply
called $L$-resolved `density of states') which can be obtained from the
previous equation as follows 
\begin{equation}
{\rm {Tr}\hat{\rho}(E;\delta)|_{L;n;p}=\int{d}r\rho_{LL}^{n;p}(r;E;\delta)}
\label{intdos}
\end{equation}

The matrix of expansion coefficients in Eq.(\ref{rhoKKR}) for a fixed cell $n
$ in layer $p$, which can be defined in terms of its components by 
\begin{equation}
\mbox{\boldmath{$\rho$}}^{n;p}(r;E;\delta)=\{\rho_{LL^{\prime}}^{n;p}(r;E;%
\delta)\} \quad ,  \label{dosmat}
\end{equation}
will be invariant under all operations of the point group. Due to the fact
that the point group is a subgroup of the rotation-reflection group 
${\cal O}(3)$
the basis functions of the IR's of the point group will be linear
combinations of spherical harmonics. Moreover it can occur that in the
truncated (l,m) basis used for calculations, which was up to $l=2$ in our
case, several spherical harmonics belong to the same IR and therefore $%
\mbox{\boldmath{$\rho$}}^{n;p}(r;E;\delta)$ will have off-diagonal elements
connecting these functions. By grouping these basis functions that belong to
the same IR $\mbox{\boldmath{$\rho$}}^{n;p}(r;E;\delta)$ becomes
block-diagonal with one block per IR. Now each block that contains more than
one basis function has to be diagonalized to find the contributions $%
\rho_j^{n;p}(r;E;\delta)$ belonging to the given IR, where $j$ labels the
eigenfunctions of $\hat{H}$. If the quantization axis is chosen normal to
the interface (${\bf z}$-direction) then at least all states $(l,m=0)$,
which are functions of ${\bf z}$ only, will belong to the identity
representation (they are invariant under all operations of the group) and
the corresponding eigenfunctions will be linear combinations of these
states. This is valid for {\it all} 2D points groups and is therefore very
general. Due to the fact that $\mbox{\boldmath{$\rho$}}^{n;p}(r;E;\delta)$
depends on $r$ and $p$ the transformation that diagonalize it will also
depend on $r$ and $p$ and therefore the basis formed by the eigenfunctions
(`eigenbasis') will gradually change in a MTJ when going from the metal into
the barrier.

In Fe(001)/Vac the underlying crystal structure is b.c.c.\ and for the (001)
surface the 2D lattice in the layers parallel to the surface is quadratic.
The point group of the 2D lattice is therefore the group $C_{4v}$
(Sch\"onflies notation) or $4mm$ (International notation) and it has 5
irreducible representations $\Delta_i,\,i=1,...,5$ where the first four are
1D and the 5th is 2D \cite{grpth}. For an angular momentum basis up to $l=2$%
, $\Delta_2$ is not realized. $\Delta_1$ is the identity representation
which is realized by $s$, $p_z$ and $d_{z^2}$ independently and we will
concentrate on the mixing of these functions.

\begin{figure}[h] 
\begin{center}
\leavevmode
\begin{tabular}{cc}
\begin{minipage}[b]{7cm}
{\hbox {\epsfxsize=7cm   
%\epsffile{/home/cu2/plots/Fe/DOS/dos-6,4eV/SPD/nobulk/dos_1_sl-0.ps}
\epsffile{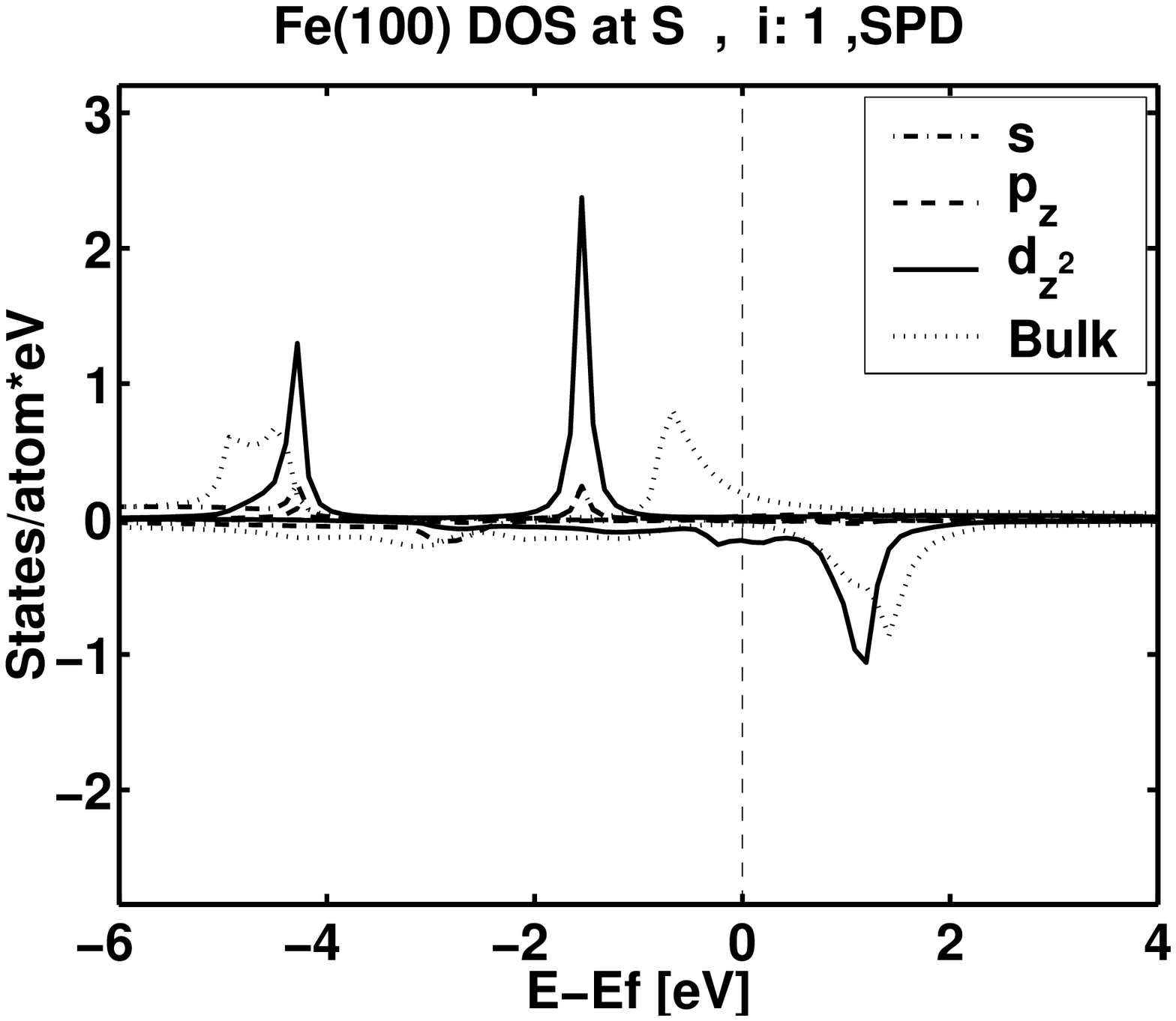}
}}
\end{minipage} &
 \begin{minipage}[b]{7cm}
 {\hbox {\epsfxsize=7cm
%\epsffile{/home/cu2/plots/Fe/DOS/dos-6,4eV/Hybrid/nobulk/dos_1_sl-0.ps}
 \epsffile{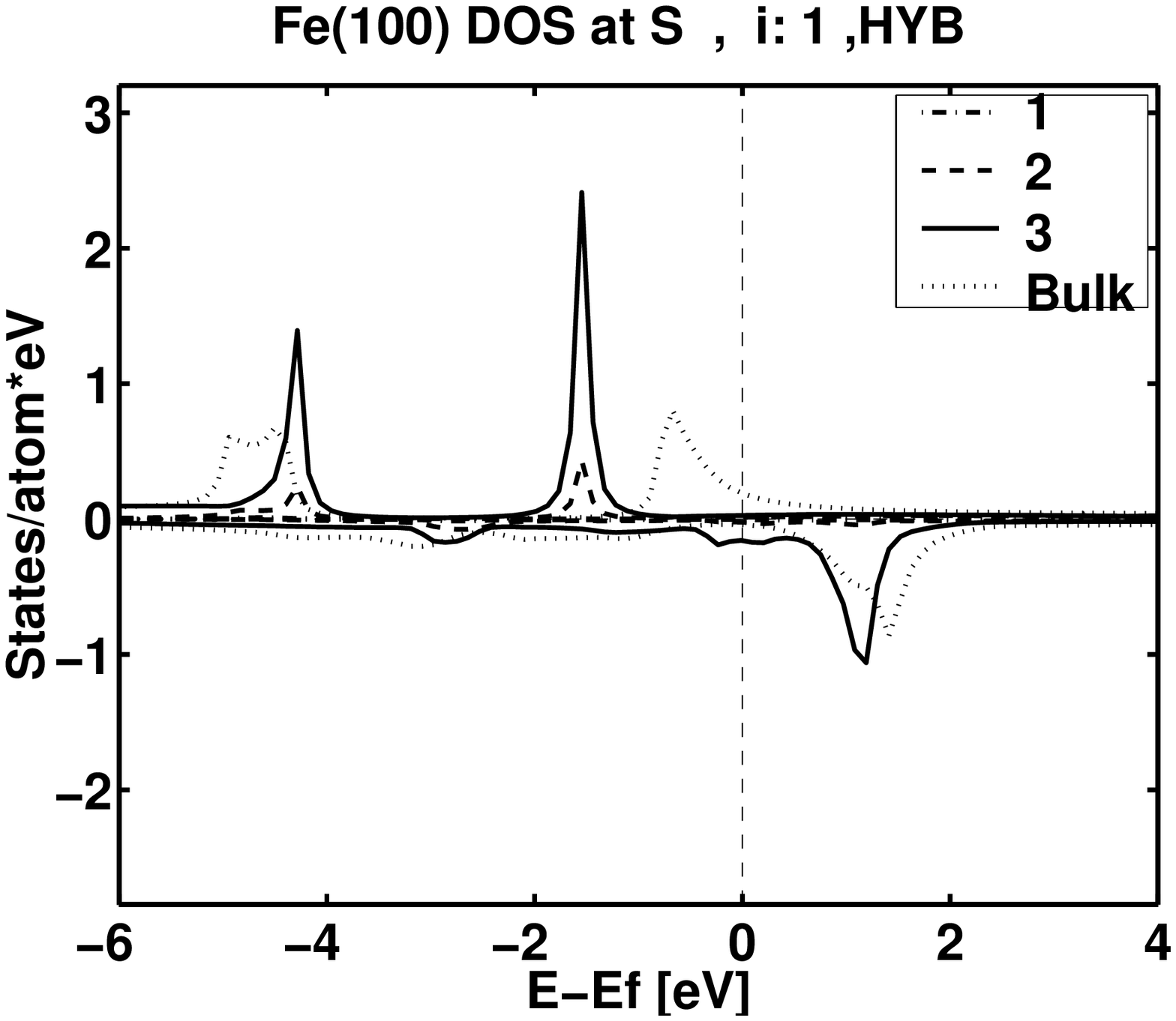}
 }}
 \end{minipage} 
 \end{tabular}
\caption{$\Delta_1$ contribution to the integrated density of states plotted 
versus energy for the surface layer of Fe(100)$|$Vac in the IR 
basis (left) and the eigenbasis (right). The minority DOS is plotted with
reversed sign.}
\label{dossl}
\end{center} 
\end{figure}

\begin{figure}[h] 
\begin{center}
\leavevmode
\begin{tabular}{cc}
 \begin{minipage}[b]{7cm}
   \epsfxsize=7cm   
%  \epsffile{/home/cu2/plots/Fe/DOS/dos-6,4eV/SPD/nobulk/dos_1_vl1.ps}
   \epsffile{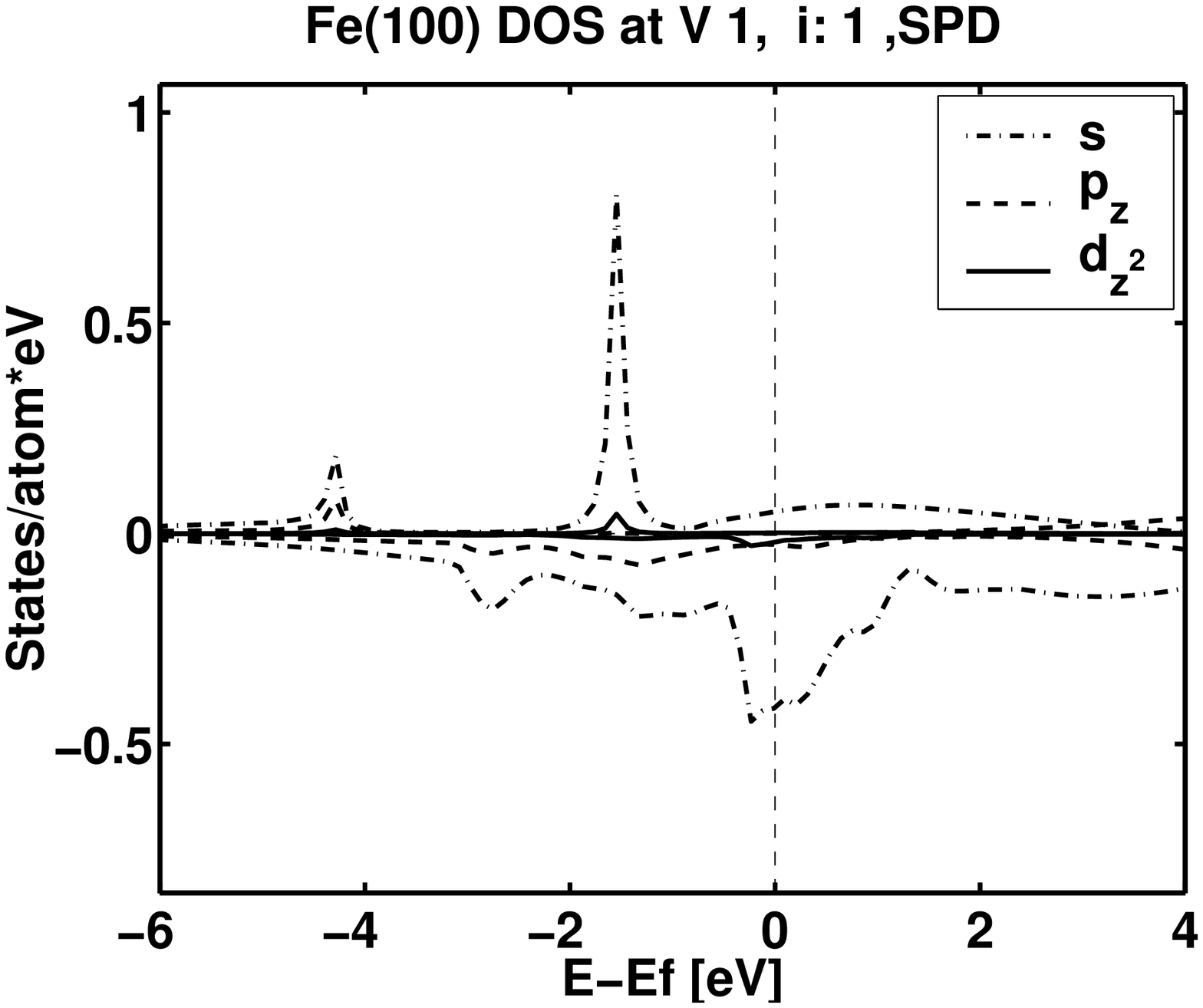}
 \end{minipage} &
 \begin{minipage}[b]{7cm}
   \epsfxsize=7cm
%  \epsffile{/home/cu2/plots/Fe/DOS/dos-6,4eV/Hybrid/nobulk/dos_1_vl1.ps}
   \epsffile{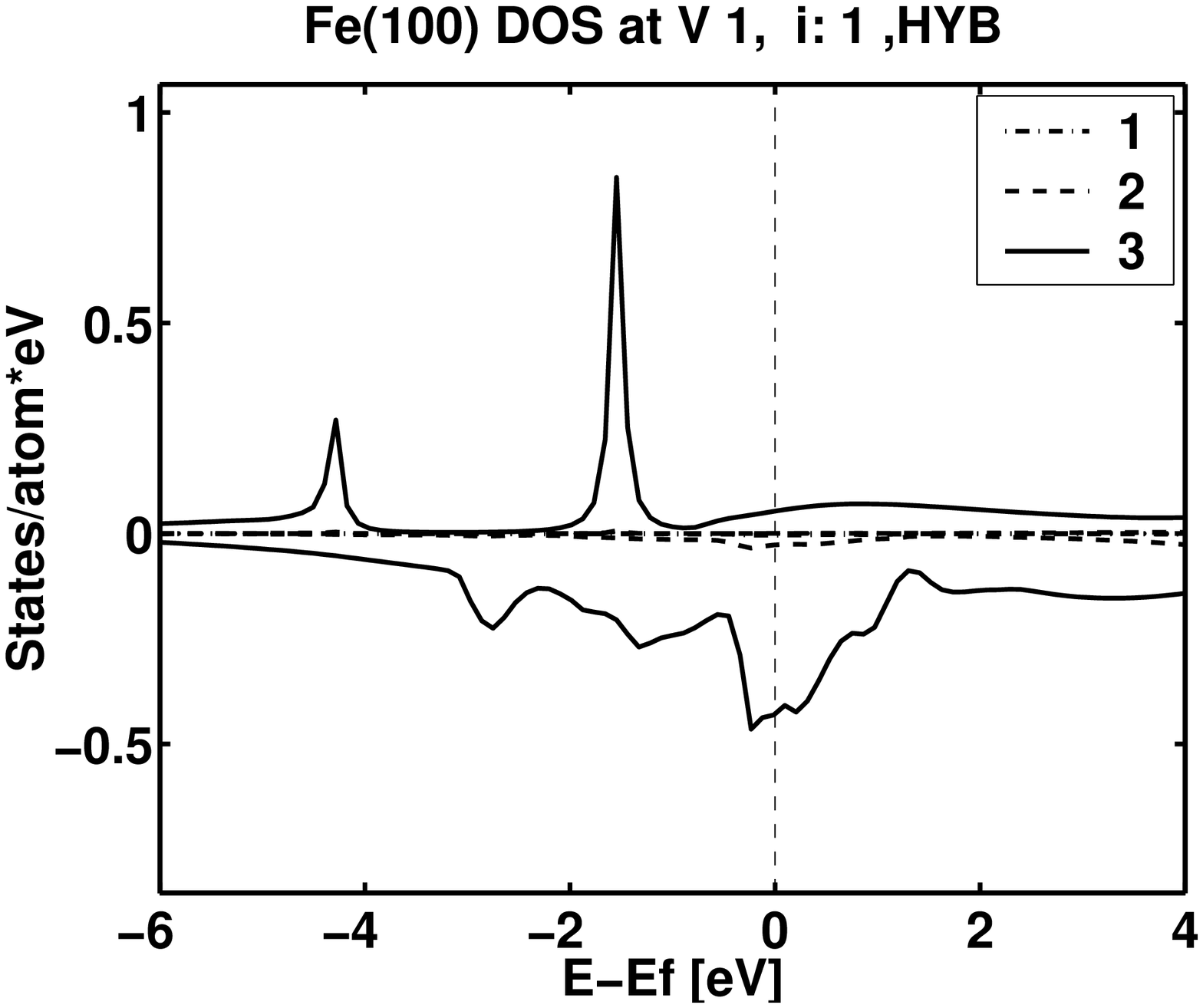}
 \end{minipage} 
\end{tabular}
\caption{$\Delta_1$ contribution to the integrated density of states plotted 
versus energy for the first vacuum layer of Fe(100)$|$Vac in the 
IR basis (left) and the eigenbasis (right). The minority DOS is plotted with 
reversed sign and 5 times enlarged.}
\label{dosv1}
\end{center} 
\end{figure}

\begin{figure}[h] 
\begin{center}
\leavevmode
\begin{tabular}{cc}
 \begin{minipage}[t]{7cm}
   \epsfxsize=7cm   
%  \epsffile{/home/cu2/plots/Fe/DOS/dos-6,4eV/SPD/nobulk/dos_1_vl2.ps}
   \epsffile{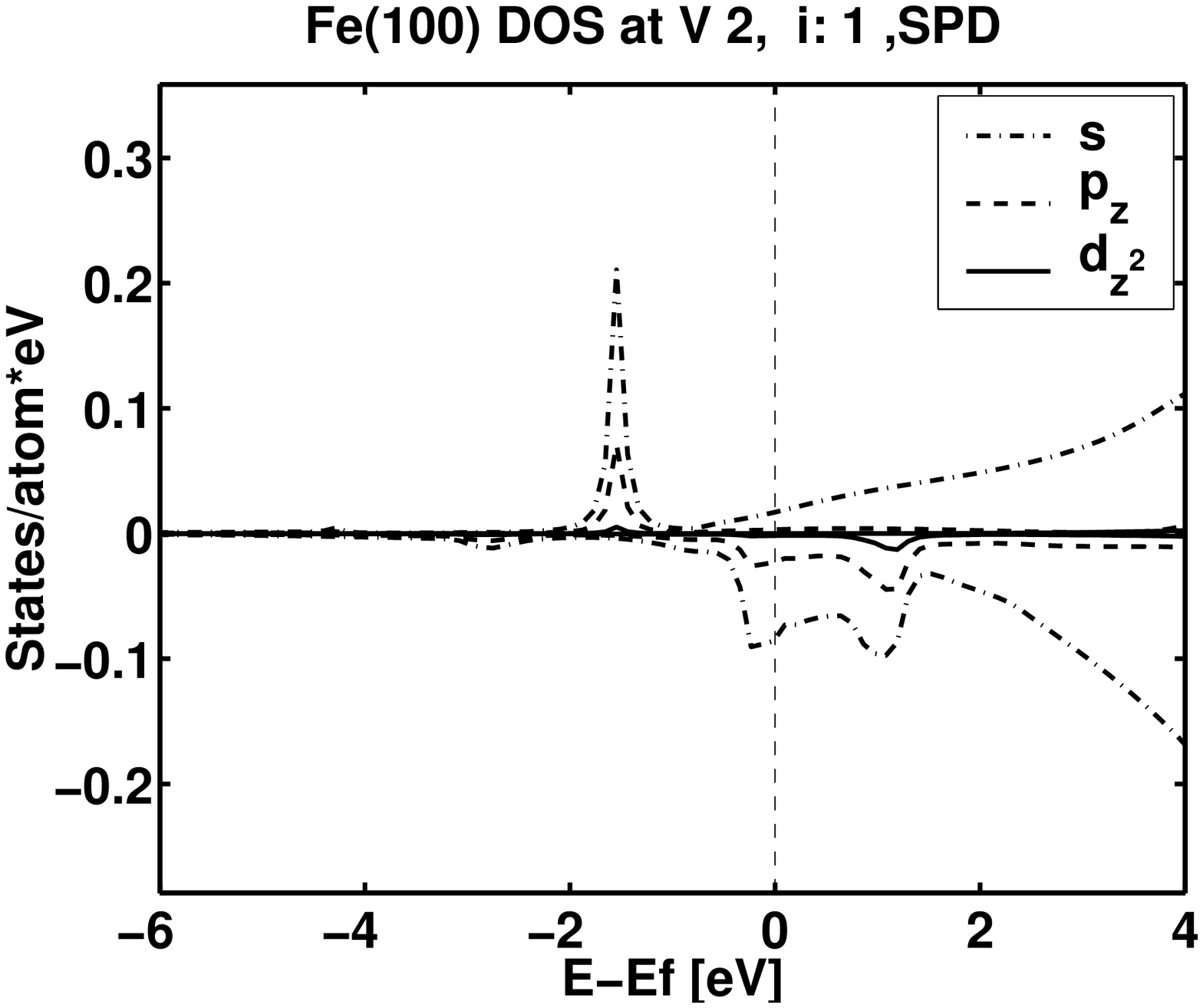}
 \end{minipage} &
 \begin{minipage}[t]{7cm}
   \epsfxsize=7cm
%  \epsffile{/home/cu2/plots/Fe/DOS/dos-6,4eV/Hybrid/nobulk/dos_1_vl2.ps}
   \epsffile{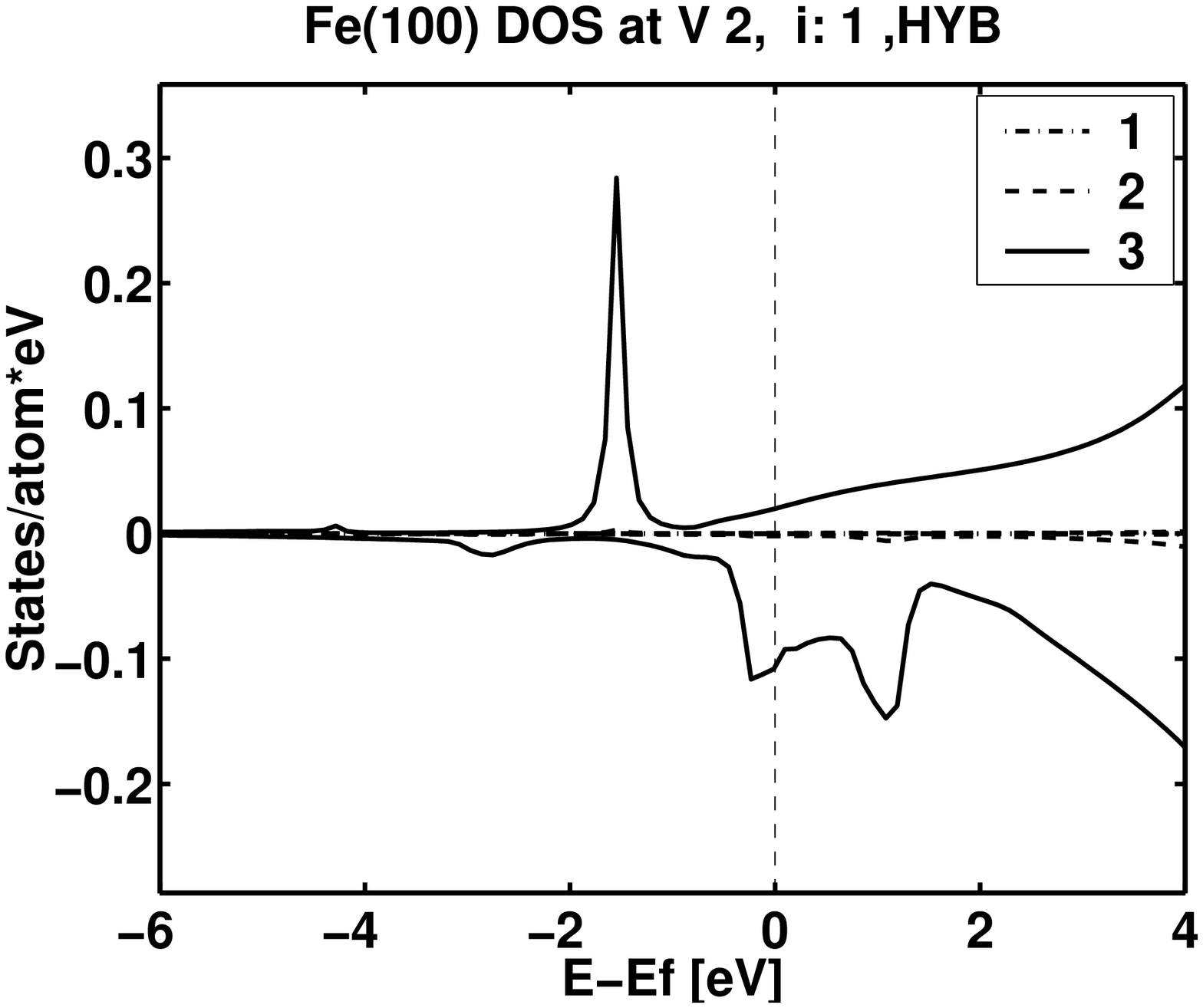}
 \end{minipage} 
\end{tabular}
\caption{$\Delta_1$ contribution to the integrated density of states plotted 
versus energy for the second vacuum layer of Fe(100)$|$Vac in the 
IR basis (left) and the eigenbasis (right). The minority DOS is plotted with
reversed sign and 5 times enlarged.}
\label{dosv2}
\end{center} 
\end{figure}

\begin{figure}[h] 
\begin{center}
\leavevmode
\begin{tabular}{cc}
 \begin{minipage}[t]{7cm}
   \epsfxsize=7cm   
%  \epsffile{/home/cu2/plots/Fe/DOS/radial/rho_1dn_spd.ps}
   \epsffile{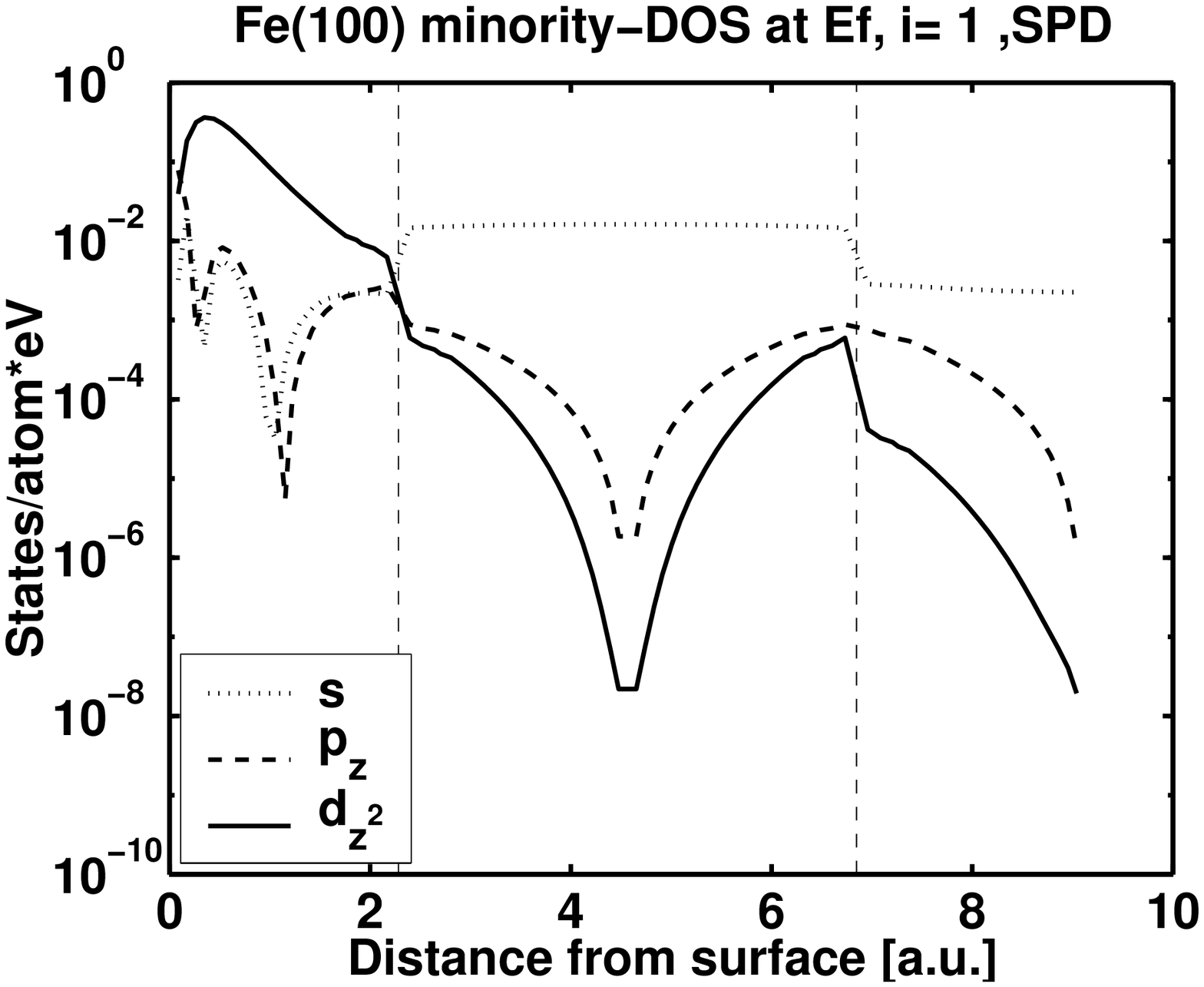}
 \end{minipage} &
 \begin{minipage}[t]{7cm}
   \epsfxsize=7cm
%  \epsffile{/home/cu2/plots/Fe/DOS/radial/rho_1dn_hyb.ps}
   \epsffile{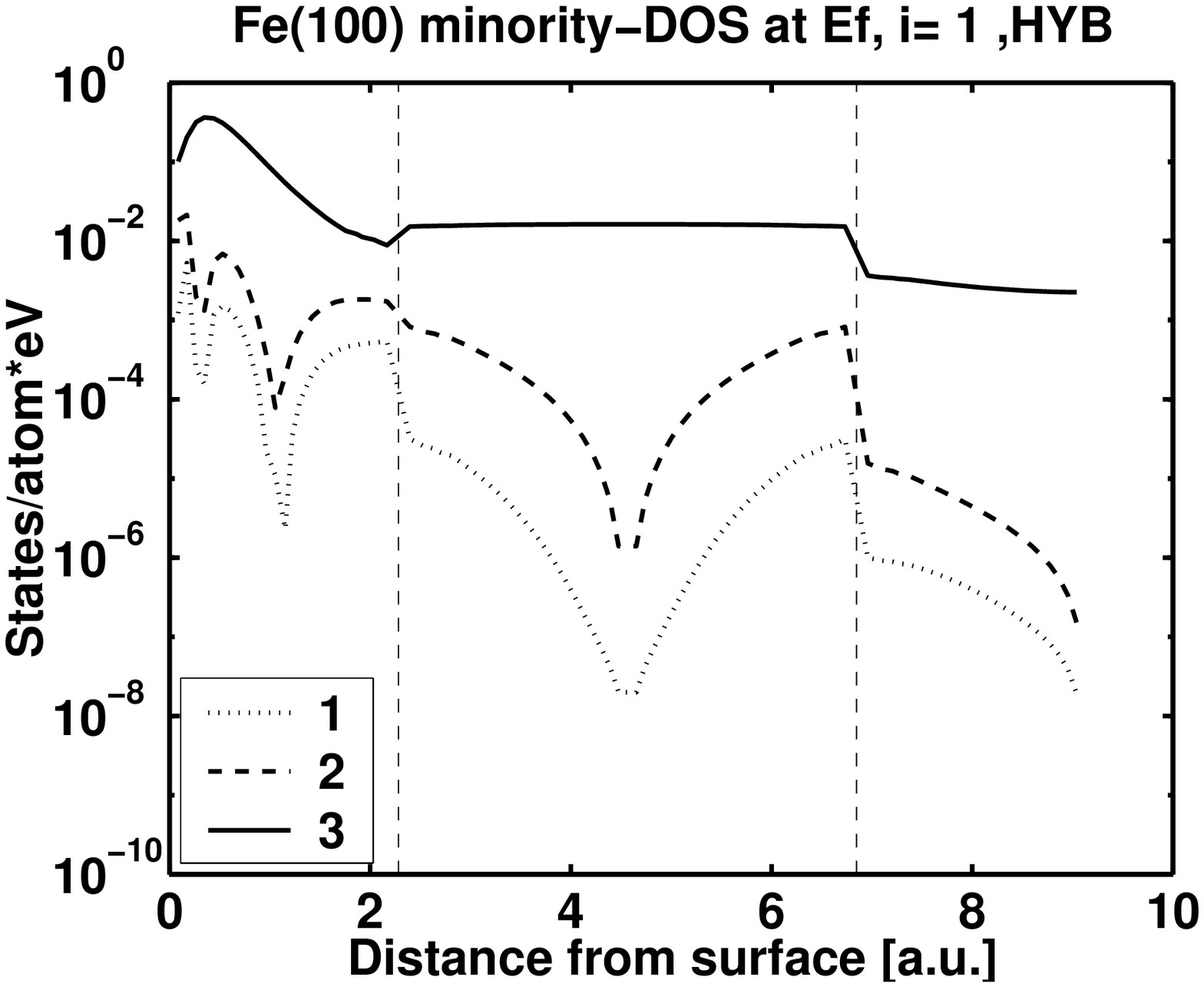}
 \end{minipage} 
 \end{tabular}
\caption{$\Delta_1$ contribution to the minority DOS matrix
$\mbox{\boldmath{$\rho$}}(r;E;\delta)$ at the Fermi Energy
plotted versus $r$ in the IR basis (left) and the eigenbasis (right). 
$r=0$ denotes the center of the cell in the surface layer 
and $r$ is measured in the direction to the vacuum.}
\label{rmin}
\end{center} 
\end{figure}

In Figs.(\ref{dossl})-(\ref{dosv2}) we plot the $L$-resolved integrated DOS
defined in Eq.(\ref{intdos}) for the surface layer and the first and second
vacuum layer against the energy for the functions belonging to $\Delta _{1}$%
. Furthermore, we plot the total $\Delta _{1}$ contribution in the bulk in
order to see if some states that occur at the surface do not occur in the
bulk and therefore may be localized. Majority contributions are plotted with
the correct sign and minority contribution with a reversed sign. The left
graph shows the DOS in the $(l,m)$ basis and the right graph shows the DOS
in the eigenbasis and we will concentrate on the $(l,m)$ basis first.
Looking at the left part of Fig.(\ref{dossl}) one can indeed see that at 1.6
eV below the Fermi Energy $E_{f}$ in the majority DOS and around $E_{f}$ in
the minority DOS there are states that have no counterpart in the bulk. To
find out if these states are localized we have a look at Fig.(\ref{dosv1})
and Fig.(\ref{dosv2}) which show the DOS in the vacuum. Indeed, one finds a
state at $E_{f}$-1.6eV in the majority DOS and around $E_{f}$ in the
minority DOS which has a similar shape as the one at the surface layer,
however we note it is largely of $s$-character whereas the state in Fig.(\ref%
{dossl}) was mainly of $d_{z^{2}}$ character. From that one could conclude
that the state found at the surface of the metal does not continue into the
vacuum and is therefore not localized. Moreover, in Fig.(\ref{dosv2}) it
seems these contributions consist of two states with $s$ and $p_{z}$
character. However, these conclusions are {\it wrong} because when we look
at the DOS in the eigenbasis (right part of Fig.(\ref{dossl})) we see that
the peaks mentioned above result from one eigenstate that extends from the
surface into the vacuum. If this were not the case there would be a state at
the surface of Fe that would be damped almost completely at the first vacuum
layer and one in the vacuum that is damped similarly at the surface of Fe,
and at some point in between Fe and vacuum the DOS from these states must
cross. This, however, is not possible because for all 2D point groups the
eigenfunctions belonging to $\Delta _{1}$ span 1D invariant spaces and
therefore degenerate eigenvalues cannot occur in the $\Delta _{1}$ block of $%
\mbox{\boldmath{$\rho$}}$; otherwise an invariant space of dimension $\geq 2$
would exist. To illustrate this fact we plotted in Fig.(\ref{rmin}) the
component resolved DOS for the minority spin channel versus $r$ by taking
the diagonal elements of $\mbox{\boldmath{$\rho$}}^{n;p}(r;E;\delta )$ in
the $(l,m)$ basis (left graph) and in the eigenbasis (right graph). $r=0$
denotes the center of a cell in the surface layer and the vertical broken
lines denotes the boundaries between layers when going from the surface into
the vacuum. One can see that when transforming from the $(l,m)$ basis to the
eigenbasis any crossings of states occurring between the surface and the
first vacuum layer are removed. The resulting eigenstates exist at the
surface {\it and} in the vacuum. While we have focused on the minority
states, the majority DOS versus $r$ behaves qualitatively in the same way.

The discontinuity of the eigenstates across the boundaries of layers in Fig.(%
\ref{rmin}) comes from the fact that in the ASA the DOS is symmetric around
the center of the cell in a layer and therefore is the same on the left and
right boundary in Fig.(\ref{rmin}). By using a full potential calculation
these discontinuities are removed, but otherwise it should qualitatively
yield the same results.

A typical wavelength of an electron at $E_{f}$ is in the order of the layer
spacing and it is only meaningful to present the average of the DOS over a
given layer, therefore we plot the integrated DOS for each eigenstate on
each layer in Fig.(\ref{dos_prof}). Besides the facts discussed in the
introduction one can see that different states within $\Delta _{1}$ have
equal decay rates in vacuum. This is also observed for states belonging to 
$\Delta _{5}$ and it follows that each IR has one decay rate which then does
not depend on the basis used. The different decay rates for the different
IR's have already been pointed out and investigated by others \cite%
{butler1,mav1}. However, when they did their analysis they only took the
trace over each block in $\int_{cell}{\rm d}r\mbox{\boldmath{$\rho$}}%
^{n;p}(r;E;\delta )$, obtaining the total contribution from each IR, and did
not analyze the functions within a block to find the true eigenfunctions of
the Hamiltonian (which are linear combinations of IR basis functions) that
change when going from the electrode into the barrier.

For completeness we want to mention that the states $d_{x^2-y^2}$ and $d_{xy}$
which belong to $\Delta_3$ and $\Delta_4$ will already be the eigenstates
of the Hamiltonian but the 2D IR $\Delta_5$ is realized by two sets of basis
functions, namely $(d_{xz},d_{yz})$ and $(p_x,p_y)$. Functions that belong
to the same row of the same IR will mix and this applies to the pairs 
$d_{xz},p_x$ and $d_{yz},p_y$. Both pairs form the same pair of eigenstates
and the two corresponding eigenvalues are twofold degenerate. The
contribution from $\Delta_3,\dots,\Delta_5$ to the surface state found 
above is about an order of magnitude smaller than from the eigenfunctions 
belonging to $\Delta_1$ near the surface and becomes negligible in large 
barriers due to their larger decay.

In summary we defined a density of states operator and its real space
representation and showed that it is very useful for analyzing the
electronic states in a magnetic tunnel junction. We showed that especially
the point group of the layers parallel to the interface between the
electrodes and the barrier and its irreducible representations can give some
insight by making the DOS matrix defined in Eq.(\ref{dosmat})
block-diagonal. But we stress that finding the eigenbasis within these
blocks cannot be provided by symmetry but has to be found by solving a
secular equation. Concentrating on the identity representation gives us a
better understanding how localized states at (001) and (111)
surfaces/interfaces in the cubic Bravais lattices can occur and that they
will be of $(s,p_{z})-d_{z^{2}}$ character {\it independent} of the actual
material used for the electrode and the barrier (the ${\bf z}$ direction is
chosen normal to the surface/interface). The general feature of $s-d_{z^{2}}$
surface states in bcc (001) surfaces of different materials found by
experiment has been mentioned in \cite{stroscio1}. Like others, see e.g. %
\cite{mav1}, they use the bulk band structure and project it onto the face
corresponding to the interface when discussing these states.
This procedure implicitly uses a fixed $(l,m)$ basis and not the eigenbasis,
therefore it cannot explain how these states are formed. Finally we want to
point out that states localized at the electrode/barrier interface play a
crucial role in tunneling, and in the case of Fe/vacuum alter one's
expectations about the spin polarization of the tunneling current. While one
would argue on the basis of the itinerant (bulk) states in the junction
that the current has a majority spin polarization, electrons can be
scattered into the localized states at the interface and thereby reverse
the spin-polarization of the tunneling current. This is reflected in the
DOS at the interface around $E_{f}$, see Fig.(\ref{dossl}) .

This work was supported by the Defense Advanced Research Projects Agency and
the Office of Naval Research (Grant No.\ N00014-96-1-1207 and Contract Nos.\
MDA972-96-C-0014, and MDA972-99-C-0009\ ). 
%\newpage 
%%%%%%%%%%%%%%%%%%%%%%%%%%%%%%%%%%%%%%%%%%%%%%%%%%%%%%%%70%%%%%%%

%\begin{references}

%%%%%%%%%%%%%%%%%%%%%%%%%%%%%%%%%%%%%%%%%%%%%%%%%%%%%%%%70%%%%%%%
%%%%%%%%%%%%%%%%%%%%%%%%%%%%%%%%%%%%%%%%%%%%%%%%%%%%%%%%70%%%%%%%
%%%%%%%%%%%%%%%%%%%%%%%%%%%%%%%%%%%%%%%%%%%%%%%%%%%%%%%%70%%%%%%%
\newpage
%%%%%%%%%%%%%%%%%%%%%%%%%%%%%%%%%%%%%%%%%%%%%%%%%%%%%%%%70%%%%%%%
%%%%%%%%%%%%%%%%%%%%%%%%%%%%%%%%%%%%%%%%%%%%%%%%%%%%%%%%70%%%%%%%
%%%%%%%%%%%%%%%%%%%%%%%%%%%%%%%%%%%%%%%%%%%%%%%%%%%%%%%%70%%%%%%%
\newpage
%%%%%%%%%%%%%%%%%%%%%%%%%%%%%%%%%%%%%%%%%%%%%%%%%%%%%%%%70%%%%%%%
%%%%%%%%%%%%%%%%%%%%%%%%%%%%%%%%%%%%

\end{document}